\documentclass[aps,prd,10pt,nofootinbib,twocolumn,eqsecnum,showpacs,superscriptaddress,preprintnumbers]{revtex4-2}

\usepackage{silence}
\usepackage[T1]{fontenc}
\usepackage[utf8]{inputenc}

\usepackage{amsmath,amssymb,amsfonts}
\usepackage{physics}

\usepackage{graphicx}
\usepackage{booktabs}
\usepackage{xcolor}
\usepackage{placeins}

\usepackage[caption=false]{subfig}

\usepackage[colorlinks=true,allcolors=blue]{hyperref}

\begin{document}

\title{Modified Cosmology from Mass-to-Horizon Relation: Background Evolution}

\author{Pranav Prasanthan}
\email{pranav.prasanthan@phd.usz.edu.pl}

\author{Hussain Gohar}
\email{hussain.gohar@usz.edu.pl}

\author{Vincenzo Salzano}
\email{vincenzo.salzano@usz.edu.pl}

\affiliation{Institute of Physics, University of Szczecin, Wielkopolska 15, 70-451 Szczecin, Poland}

\date{\today}

\begin{abstract}
We investigate the cosmological implications of the mass-to-horizon relation, which provides a unified framework for thermodynamically consistent generalized horizon-entropy functionals. Using the Cai--Kim formulation of the first law of thermodynamics, we derive the corresponding modified Friedmann equations and examine the resulting background evolution. We find that cosmological viability sharply restricts admissible deviations from the Bekenstein--Hawking area law: phenomenologically acceptable scenarios are confined to a narrow neighborhood of the standard entropy, while more pronounced deviations generically spoil the standard radiation--matter--dark-energy sequence. Power-law entanglement corrections can give rise to a moderate early-dark-energy component, but only within a tightly constrained region of parameter space, whereas quantum-gravity corrections are suppressed by the Planck scale and remain observationally irrelevant. Consequently, all viable models predict a $\Lambda$CDM-like cosmological background at the present epoch. These findings demonstrate that background cosmology alone imposes stringent constraints on thermodynamically consistent generalized entropy constructions of this class.
\end{abstract}
\maketitle

\section{Introduction}
\label{sec:introduction}

A wide range of cosmological observations indicates that the Universe is undergoing a phase of accelerated expansion \cite{SupernovaCosmologyProject:1998vns, SupernovaSearchTeam:1998fmf,WMAP:2012nax,Planck:2018vyg,eBOSS:2020yzd,Brout:2022vxf,DES:2022ccp,DESI:2025zgx}. Within General Relativity (GR), this phenomenon is commonly attributed to dark energy, often modeled as a cosmological constant $\Lambda$ with equation of state $\omega=-1$. 
Although the resulting $\Lambda$ cold dark matter ($\Lambda$CDM) model successfully describes a broad range of observations \cite{Amendola:2015ksp}, it faces well-known theoretical challenges, including the fine-tuning and coincidence problems \cite{Weinberg:1988cp,Sahni:1999gb,Carroll:2000fy,Padmanabhan:2002ji}, as well as observational tensions such as the $H_0$ and $S_8$ discrepancies \cite{CosmoVerseNetwork:2025alb}. These issues have motivated the exploration of alternative cosmological scenarios based on modified gravity and dynamical dark energy \cite{CANTATA:2021asi,Li:2011sd,Copeland:2006wr}.

An alternative route is provided by the gravity--thermodynamics correspondence, which relates gravitational dynamics to horizon thermodynamics \cite{Cai:2005ra}. Following the pioneering works of Bekenstein and Hawking \cite{Bekenstein:1973ur,Hawking:1974rv,Hawking:1975vcx}, Jacobson showed that Einstein's equations can be derived from the Clausius relation applied to local causal horizons \cite{Jacobson:1995ab}. This thermodynamic perspective was subsequently extended to Friedmann--Lemaître--Robertson--Walker (FLRW) spacetimes with any spatial curvature by Cai and Kim \cite{Cai:2005ra}, who derived the Friedmann equations from the first law of thermodynamics applied to the apparent horizon. Within this framework, the standard Bekenstein--Hawking entropy reproduces the standard $\Lambda$CDM equations, while modifications of the entropy-area relation generate modified Friedmann equations; the resulting corrections can be reinterpreted as an effective dark energy sector \cite{Lymperis:2018iuz}.

The Bekenstein definition of horizon entropy is intrinsically nonextensive and nonadditive, as it scales with the area of the horizon rather than with the enclosed volume \cite{Hawking:1971tu,Bekenstein:1972tm,Gibbons:1977mu}. This nonextensive character has motivated the development and systematic investigation of generalized entropy frameworks—such as Tsallis, Rényi, Sharma–Mittal, Kaniadakis, Tsallis–Cirto, Tsallis–Zamora, and Barrow entropies \cite{Tsallis:1987eu,Renyi:1959pbs,sharma1975new,sharma1977new,Kaniadakis:2002zz,Kaniadakis:2005zk,Tsallis:2012js,Zamora:2022cqz,Barrow:2020tzx}—which have been extensively employed in gravitational and cosmological settings \cite{Nojiri:2022aof,Nojiri:2022sfd,Promsiri:2020jga,Promsiri:2021hhv,Tannukij:2020njz,Cimidiker:2023kle,Nakarachinda:2021jxd,Czinner:2025koi,Nakarachinda:2022gsb,Saridakis:2020zol,Dabrowski:2020atl,Komatsu:2015nkb,Komatsu:2016vof,Liu:2022snq,Majhi:2017zao,Luciano:2021mto,DiGennaro:2022ykp,DiGennaro:2022grw,Asghari:2021bqa,Abreu:2022pil}. In addition, quantum-gravitational and entanglement-induced corrections to the Bekenstein–Hawking entropy, which yield logarithmic and power-law modifications to the Bekenstein entropy, have also been the subject of extensive analysis \cite{Rovelli:1996dv,Medved:2004yu,Medved:2004eh,Das:2007mj,Alonso-Serrano:2018ycq,Alonso-Serrano:2020hpb}.

Recent studies have demonstrated that employing generalized entropies on black hole and cosmological horizons in conjunction with the Hawking temperature leads to thermodynamic and holographic inconsistencies \cite{Nojiri:2021czz,Cimdiker:2022ics,Gohar:2023hnb,Gohar:2023lta,Gohar:2025yfx,Gohar:2026hiy}. To restore consistency, the notion of a mass-to-horizon relation was introduced in \cite{Gohar:2023lta} to maintain holographic thermodynamic consistency, resulting in a generalized entropy functional that encompasses several non-extensive and phenomenological entropy frameworks. More recently, this construction has been extended to incorporate quantum-gravity and entanglement corrections, thereby providing a unified treatment of both non-extensive and quantum-gravity/entanglement-corrected entropies while preserving the Clausius relation and ensuring holographic thermodynamic consistency \cite{Gohar:2025yfx,Gohar:2026hiy}.

In this work, we investigate the cosmological consequences of this generalized mass-to-horizon entropy in \cite{Gohar:2025yfx}. Applying the Cai--Kim formulation \cite{Cai:2005ra} of the first law of thermodynamics to the apparent horizon, we derive the corresponding modified Friedmann equations and reinterpret the resulting corrections as an effective dark energy component. We then solve the cosmological equations numerically and study the viability of the model across the parameter space. We find that the background evolution is extremely sensitive to departures from the standard area law, strongly constraining the entropy parameters and confining viable cosmologies to a narrow region close to the Bekenstein--Hawking limit.

The remainder of the manuscript is organized as follows. In Sec.~\ref{sec:First}, we briefly review the thermodynamic description of the apparent horizon in a FLRW spacetime and formulate the first law of thermodynamics on the apparent horizon and recover the standard Friedmann equations within the context of GR. In Sec.~\ref{sec:3}, we introduce the generalized mass-to-horizon entropy, derive the corresponding modified Friedmann equations, and present the dimensionless formulation and parameter cases. In Sec.~\ref{Discussion}, we present the resulting background evolution for each case and discuss the physical implications, with emphasis on observationally and theoretically motivated parameter ranges. We summarize our main conclusions in Sec.~\ref{conclusion}.

\section{Thermodynamics of the apparent horizon}
\label{sec:First}
\subsection{\texorpdfstring{$\rm FLRW$}{FLRW} metric and the apparent horizon}
We consider a spatially homogeneous and isotropic universe governed by FLRW metric. In the coordinates $(ct, r, \theta, \phi)$, the line element is given by \cite{Bak:1999hd,Cai:2005ra},
\begin{equation}
    ds^2 = -c^2dt^2 + a(t)^2\left[\frac{dr^2}{1-kr^2}+r^2d\Omega^2\right],
\end{equation}
where $(r, \theta, \phi)$ are co-moving spatial coordinates, $a(t)$ is the scale factor as a function of cosmic time $t$, and $k = -1, 0, 1$ corresponds to an open, flat, or closed universe, respectively. Here, $d\Omega^2 = d\theta^2 + \sin^2\theta \, d\phi^2$ denotes the metric on the unit two-sphere, and $c$ is the speed of light. By using the spherical symmetry, the above metric can be rewritten as \cite{Bak:1999hd}
\begin{equation}
    ds^2 = h_{\alpha \beta}dx^{\alpha}dx^{\beta} + r_{\text{phys}}^2(d\theta^2 +\sin^2\theta d\phi^2),
\end{equation}
where $r_{\text{phys}} = a(t)r$ is the physical radius and the two-dimensional metric in the ($x^0=ct, x^1=r$) plane takes the form
\begin{equation}\label{metric}
    h_{\alpha\beta} = \text{diag}\left[-1, \frac{a(t)^2}{1-kr^2}\right].
\end{equation}
We model the matter-energy content of the FLRW universe using the perfect-fluid energy-momentum tensor $T_{\mu\nu}$, defined as \cite{steiner2007solution}
\begin{equation}
    T_{\mu\nu} = \left(\varepsilon + P\right)\frac{U_{\mu}U_{\nu}}{c^2} + P g_{\mu\nu}
\end{equation}
where $\varepsilon$ and $P$ are the energy density and pressure of the fluid, $U_{\mu}$ is the four-velocity of the fluid, and $g_{\mu\nu}$ is the metric tensor, respectively. 
The projection of a $(3+1)$-dimensional energy-momentum tensor $T^{\mu\nu}$ onto this two-dimensional space normal to the two-spheres leads to two invariant quantities, the scalar work density $W$ and the energy-supply co-vector $\psi_a$ \cite{Hayward:1997jp,Bak:1999hd}, which for a perfect fluid takes the form \cite{Cai:2006rs,Tian:2014sca}
\begin{equation}
    W = \frac{\varepsilon-P}{2}, \quad  \psi = \psi_{\text{t}}+\psi_{\text{r}},
\end{equation} 
where $\psi_{\text{t}}$ and $\psi_{\text{r}}$ are the time 
and radial components of the energy-supply vector, 
describing the energy flux across the horizon. 

In order to investigate the thermodynamic quantities, we use the apparent horizon for the FLRW universe, which is a quasi-local and dynamical boundary that always exists in FLRW spacetime, independently of the spatial curvature, and provides a natural causal boundary for a comoving observer \cite{Faraoni:2015ula}. The apparent horizon can be endowed with gravitational entropy, surface gravity, and the Cai-Kim temperature, and admits consistent formulations of the first and second laws of thermodynamics \cite{Hayward:1997jp,Hayward:1998ee,Bak:1999hd,Wang:2005pk,Cai:2008gw}. For this reason, it is widely regarded as the appropriate horizon on which to formulate the thermodynamics of the Universe. The region enclosed by the apparent horizon may therefore be treated as an open thermodynamic system \cite{Tian:2014sca}, whose internal energy changes due to the flux of matter-energy crossing the horizon as a consequence of cosmic expansion. Hence, the apparent horizon is defined by the condition \(h^{\alpha\beta} \partial_{\alpha} r_{\text{phys}} \, \partial_{\beta} r_{\text{phys}} = 0\), which implies that the vector \(\partial_{\alpha} r_{\text{phys}}\) is null. Evaluating this condition using the metric in Eq. \eqref{metric} yields the mathematical expression for the apparent horizon radius $r_a$ as
\begin{equation}\label{appa}
    r_a = \frac{c}{\sqrt{H^2 + \frac{k c^2}{a^2}}},
\end{equation}
where $H = \frac{\dot a}{a}$ is the Hubble parameter with dots representing derivatives with respect to cosmic time $t$ and $h^{\alpha\beta}$ is the inverse of the metric in Eq.\eqref{metric}. Note that for a spatially flat FLRW universe $(k=0)$, the apparent horizon coincides with the Hubble horizon $r_H = \frac{c}{H}$.

\subsection{Horizon thermodynamics}

The apparent horizon can be endowed with thermodynamic properties through the holographic principle \cite{tHooft:1993dmi,Susskind:1994vu,Bak:1999hd}. Within General Relativity, its entropy is given by the Bekenstein--Hawking relation \cite{Bekenstein:1973ur,Hawking:1974rv,Hawking:1975vcx}
\begin{equation}\label{bekenstein entropy}
S_{\text{BH}}=\frac{k_B c^3}{4\hbar G}A,
\end{equation}
where $A=4\pi r_a^2$ is the apparent-horizon area.

Two temperature definitions are commonly considered. The first is the Cai--Kim temperature \cite{Cai:2005ra},
\begin{equation}\label{cai}
T_{\text{CK}}=\frac{\hbar c}{2\pi k_B r_a},
\end{equation}
which corresponds to the Hawking-like temperature associated with the apparent horizon \cite{Cai:2008gw}. The second is the Kodama--Hayward temperature \cite{Cai:2006rs}. Because it can be non-positive, its partial absolute value is used \cite{Tian:2014sca}, which reads
\begin{equation}\label{Th+}
T_{\text{H}}^+=\frac{\hbar c}{2\pi k_B r_a}
\left(1-\frac{\dot r_a}{2Hr_a}\right),
\end{equation}
where positivity requires the additional condition $\frac{\dot r_a}{2Hr_a}<1$ \cite{Sheykhi:2008qr}.
As shown by Tian and Booth \cite{Tian:2014sca}, this condition implies that $T_{\text{H}}^+$ vanishes in the radiation-dominated era ($\omega=1/3$) and becomes negative for $\omega>1/3$, rendering it thermodynamically problematic. Taking instead the  absolute value avoids the negative regime but still vanishes at $\omega=1/3$. By contrast, the Cai--Kim temperature remains positive throughout cosmic evolution and is consistent with the Clausius relation
\begin{equation}
dE=c^2dM=T_hdS_h, \label{fun}
\end{equation}
implying the equivalence of mass and energy associated with the apparent horizon. More importantly, it is justified by quantum field theory \cite{Cai:2008gw}. For these reasons, we adopt the Cai--Kim temperature as the associated temperature of the apparent horizon throughout this work.

\subsection{First law of thermodynamics on the apparent horizon}
\label{sec:firstlaw}

Following the Cai--Kim formulation \cite{Cai:2005ra}, the first law of thermodynamics on the apparent horizon reads as
\begin{equation}\label{fi}
\delta Q=-dE=T_h dS_h,
\end{equation}
where $\delta Q$ is the heat (energy) flow across the apparent horizon, $T_h$ and $S_h$ denote the horizon temperature and entropy, respectively. Within this formulation, in an { infinitesimal isochoric process ($\dot r_a=0$)}, the energy change inside the horizon is entirely due to the flux of the cosmic fluid across the apparent horizon, which is defined as in terms of time component of energy flow vector \cite{Cai:2005ra,Tian:2014sca}, such that
\begin{equation}\label{flux}
\delta Q=-dE= -A\psi_{\text{t}}=
4\pi r_a^3H(\varepsilon+P)dt,
\end{equation}
with $\varepsilon$ and $P$ the total energy density and pressure.

Substituting the Bekenstein--Hawking entropy \eqref{bekenstein entropy}, the Cai--Kim temperature \eqref{cai}, and the geometric relation
\begin{equation}\label{eq:RA_dot_geom}
\dot r_a=-\frac{r_a^3H}{c^2}
\left(\dot H-\frac{kc^2}{a^2}\right),
\end{equation}
into Eq.~\eqref{fi} yields
\begin{equation}\label{eq:acceleration_equation}
-\frac{4\pi G}{c^2}(\varepsilon+P)
=
\dot H-\frac{kc^2}{a^2}.
\end{equation}
Combining this result with the continuity equation
\begin{equation}\label{continuity equation}
\dot\varepsilon+3H(\varepsilon+P)=0,
\end{equation}
and integrating, one obtains
\begin{equation}
H^2+\frac{kc^2}{a^2}=
\frac{8\pi G}{3c^2}\varepsilon
+C,
\end{equation}
where $C$ is an integration constant with dimension $[\text{T}]^{-2}$. This constant emerges from the energy flow and associated entropy change at the apparent horizon, effectively acting as a constant energy density that drives the late-time acceleration equivalent to a cosmological constant. Therefore, choosing  $C\equiv\frac{\Lambda c^2}{3}$, we recover 
\begin{equation}\label{LCDM COSM}
H^2+\frac{kc^2}{a^2}=
\frac{8\pi G}{3c^2}\varepsilon
+\frac{\Lambda c^2}{3},
\end{equation}
namely the standard Friedmann equation of $\Lambda$CDM.

Within the Cai-Kim framework, any modification of the horizon entropy induces corresponding corrections to the Friedmann equations, which can be interpreted as an effective dark-energy sector \cite{Lymperis:2018iuz}. In the following section, we replace the Bekenstein--Hawking entropy by the generalized mass-to-horizon entropy and derive the resulting modified cosmological dynamics.

\section{Mass-to-Horizon entropy and modified cosmology}
\label{sec:3}
Within holographic frameworks applied to black hole thermodynamics and cosmology, thermodynamic consistency constitutes a fundamental requirement. In this context, thermodynamic consistency refers to the self-consistent definition of horizon thermodynamic quantities, namely the entropy, temperature, and associated mass (or energy), such that they satisfy the Clausius relation. This condition ensures that the mass and energy assigned to the horizon are thermodynamically equivalent for the chosen entropy and temperature.

Recent investigations have demonstrated that generalized horizon entropy formalisms, when combined with the Cai--Kim temperature or Hawking temperature, generally fail to satisfy the Clausius relation \cite{Nojiri:2021czz,Cimdiker:2022ics} unless an appropriate mass--to--horizon relation is introduced \cite{Gohar:2023hnb,Gohar:2023lta}. The MHR provides the additional thermodynamic structure required to restore equilibrium consistency and establish a well-defined correspondence between horizon thermodynamics and gravitational dynamics.

For the standard Bekenstein--Hawking entropy and Cai--Kim temperature, thermodynamic consistency is achieved through the linear MHR
\begin{equation}
M=\frac{c^2}{G}r_a,
\end{equation}
which was first proposed in Ref.~\cite{Gohar:2023hnb}. This framework was subsequently generalized to encompass a broad class of horizon entropies, including nonextensive, quantum-gravitational, entanglement-induced, and phenomenological corrections \cite{Gohar:2023lta,Gohar:2025yfx,Gohar:2026hiy}. The resulting generalized MHR can be expressed as
\begin{equation}\label{EMHR}
M = \gamma \frac{c^2}{G}\ell_p
\left[
\frac{r_a}{\ell_p}
\mp
\beta
\left(\frac{r_a}{\ell_p}\right)^{3-\alpha}
\right]^m ,
\end{equation}
where $\gamma$ and $\beta$ are dimensionless parameters, $m$ and $\alpha$ characterize the scaling behavior of the generalized relation, and $\ell_p$ denotes the Planck length. The above expression reduces to the linear MHR in the appropriate parameter limits and provides a unified description of a wide range of generalized horizon thermodynamic scenarios.

Combining Eq.~\eqref{EMHR} with the Cai--Kim temperature through Eq.~\eqref{fun} and assuming $\beta(r_a/\ell_p)^{2-\alpha}\ll1$, it yields the generalized mass-to-horizon entropy \cite{Gohar:2025yfx}
\begin{equation}\label{eq:entropy_def}
S_G=
2\pi\gamma k_B
\left[
\frac{m}{m+1}
\left(\frac{r_a}{\ell_p}\right)^{m+1}
\mp
m\beta
\frac{\sigma-1}{\sigma}
\left(\frac{r_a}{\ell_p}\right)^\sigma
\right],
\end{equation}
where $\sigma=m+3-\alpha$. This construction provides a unified thermodynamic framework encompassing several well-known generalized entropy formalisms as particular limits. The standard Bekenstein--Hawking entropy is recovered for $m=\gamma=1$ and $\beta=0$. Logarithmic corrections arising from quantum-gravitational effects are obtained in the limit $m=1$ and $\alpha\rightarrow 4$ (equivalently, $\sigma\rightarrow 0$) \cite{Rovelli:1996dv,Medved:2004yu,Medved:2004eh,Alonso-Serrano:2018ycq}, while entanglement-induced corrections emerge for $m=\beta=1$ with $\alpha$ treated as a free parameter \cite{Das:2007mj}. Setting $\beta=0$ and choosing $m=2\delta-1$ reproduces the Tsallis--Cirto entropy \cite{Tsallis:2012js}, whereas the choice $m=1+\Delta$, with $0\leq\Delta\leq1$, yields the Barrow entropy \cite{Barrow:2020tzx}.

{ The generalized MHR is physically mandated by the requirement of holographic thermodynamic consistency \cite{Gohar:2023hnb, Gohar:2026hiy, Gohar:2025yfx,Cimdiker:2022ics, Nojiri:2021czz, Gohar:2023lta}. For a cosmological apparent horizon, the Cai--Kim temperature is determined by the surface gravity and justified by quantum field theory (QFT) \cite{Cai:2008gw}, yielding $T_h=\hbar c/2\pi k_B r_a$ \eqref{cai}, and any ad hoc modification of this expression is not supported by QFT considerations. Nevertheless, for a generic entropy functional $S_G(r_a)$, the product $T_h\,dS_G$ does not, in general, constitute an exact differential; the integrability condition $c^2dM/dr_a = T_h\,dS_G/dr_a$ is violated unless the mass function $M(r_a)$ is suitably generalized. The proposed MHR is explicitly constructed to satisfy this integrability requirement, thereby ensuring that the Clausius relation integrates identically to $E = Mc^2$ without necessitating any modification of $T_h$. 

From a geometric perspective~\cite{Gohar:2026hiy}, this construction is associated with the Misner--Sharp quasilocal energy and the Wald entropy functional, and is justified via the Jacobson approach, wherein the effective gravitational coupling obeys $G_{\text{eff}}^{-1}\propto dM/dr_a$, thus encoding the entropy parameters $(m,\beta,\alpha)$ directly into the scale dependence of $G_{\text{eff}}$ in scalar--tensor theories of gravity. Furthermore, it has been implemented within the Iyer--Wald approach in the context of $f(R)$ gravity in \cite{Mondal:2026mqb}. Consequently, the MHR furnishes a thermodynamically consistent and geometrically well-defined framework that preserves the QFT origin of $T_h$ while reproducing, in the appropriate limits, the Bekenstein--Hawking, quantum-corrected, Tsallis--Cirto, and Barrow entropy functions.
}

{ Moreover, physical consistency requires that the
generalized mass-to-horizon relation remain non-negative
and the associated entropy remain strictly positive. Since, by construction, $m>0$ and $\gamma>0$, the positivity conditions are predominantly controlled by the parameters $\alpha$ and $\beta$. For the upper ($-$) branch, the correction term in MHR \eqref{EMHR} must remain subdominant, which leads to the constraint
\begin{equation}
1-\beta\left(\frac{r_a}{\ell_p}\right)^{2-\alpha}\ge0,
\end{equation}
supplemented by the entropy-positivity requirement
\begin{equation}
\frac{m}{m+1}\left(\frac{r_a}{\ell_p}\right)^{m+1}
>
m\beta\frac{\sigma-1}{\sigma}
\left(\frac{r_a}{\ell_p}\right)^{\sigma},
% \qquad
% \sigma=m+3-\alpha.
\end{equation}
In particular, for cosmological horizons ($r_a/\ell_p\gg1$), the first inequality is automatically satisfied for $\alpha>2$. For $\alpha = 2$, non-negativity requires $\beta\le1$, whereas for $\alpha<2$ it imposes an upper bound on the admissible values of $\beta$. For the lower ($+$) branch, the mass $M$ is explicitly positive for $\beta>0$, while the entropy remains positive provided $\sigma>1$, which is equivalent to the condition $\alpha<m+2$.

An important physical implication of the generalized MHR is the regularization of the temperature divergence that affects the standard scenario. For $\beta = 0$, the mass scales as $M \propto r_a^m$, while the Cai–Kim temperature behaves as $T_h \propto r_a^{-1}$, leading to the limit $T_h \to \infty$ as $M \to 0$. In contrast, for $\beta \neq 0$ and $\alpha>2$, introducing $x \equiv r_a/\ell_p$ and $M_0 \equiv \gamma(c^2/G)\,\ell_p$, the minus-sign branch of the MHR
\[
M_-(x) = M_0 x^m \bigl[1 - \beta x^{2-\alpha}\bigr]^m,
\]
exhibits a minimum horizon radius 
\[
x_{\min} = \beta^{-1/(2-\alpha)},
\]
 below which $M$ becomes negative and is therefore excluded on physical grounds. At this minimum radius, the mass vanishes, $M_{\min}=0$, while the entropy remains finite. Using the condition $\beta x_{\min}^{2-\alpha} = 1$, which implies $\beta x_{\min}^{\sigma} = x_{\min}^{m+1}$ with $\sigma = (m+1) + (2-\alpha)$, the second term in brackets simplifies, yielding
\[
S^{(-)}_{G,\min}
= 2\pi \gamma k_B \, m \, x_{\min}^{m+1}
\left[ \frac{1}{m+1} - \frac{\sigma-1}{\sigma} \right],
\]
where $x_{\min}^{m+1} = \beta^{-(m+1)/(2-\alpha)}$. This entropy is positive provided $0<\sigma<\tfrac{m+1}{m}$, which coincides with the subdominance (positivity) conditions imposed previously; under these conditions one has $S^{(-)}_{G,\min}>0$. Consequently, the generalized framework introduces a natural cutoff that prevents the temperature from attaining the divergent limit $T_h \to \infty$, thereby guaranteeing that the thermodynamic description remains physically well-defined for all admissible horizon radii.

The existence of a nonvanishing minimum entropy at zero mass arises directly from the deformation parameters $(\beta,\alpha,m)$ and constitutes a characteristic signature of the underlying quantum-gravitational or entanglement-driven modifications to the horizon structure. In the particular case of the logarithmic correction, $m = 1$ and $\alpha \to 4$ (i.e., $\sigma \to 0$), the entropy \eqref{eq:entropy_def}
reduces to
\[
S^{(-)}_G(x)
=
\pi\gamma k_Bx^2
+
2\pi\gamma k_B\beta\ln x
+
C,
\]
up to an additive constant $C$. The corresponding minimum horizon radius $x_{\min} = \sqrt{\beta}$ then yields
\[
S^{(-)}_{G,\min}
=
\pi\gamma k_B\beta(1+\ln\beta)+C.
\]
This behavior supports the consistency of the MHR construction and agrees with the qualitative predictions obtained from quantum-gravity-induced corrections to the Bekenstein entropy \cite{Adler:2001vs,Chen:2014jwq,Ong:2018syk,Cimdiker:2022ics,Alonso-Serrano:2020hpb}. For the plus-sign branch, \[M_+(x) = M_0 x^m [1 + \beta x^{2-\alpha}]^m\] with \(\beta>0\), the bracketed factor is strictly positive for all \(x>0\). Consequently, no finite minimum radius exists. The corresponding entropy,
\[
S_G^{(+)}(x) = 2\pi\gamma k_B \left[\frac{m}{m+1}x^{m+1} + \frac{m\beta(\sigma-1)}{\sigma}\,
x^\sigma\right],
\]
therefore does not regularize the divergence of the temperature. 

In the logarithmic limit $m=1$, $\alpha\to4$, one obtains
\[
S^{(+)}_G(x)
=
\pi\gamma k_Bx^2
-
2\pi\gamma k_B\beta\ln x
+
C,
\]
which diverges as $x\to0$, while the mass behaves as \(M_+(x) \sim M_0 \beta/x\). More generally, for \(x\to 0\) and \(\alpha>2\), the bracketed term dominates, and the mass scales as
\[
M_+(x) \sim M_0 \beta^m x^{m(3-\alpha)}.
\]
This expression diverges only for \(\alpha>3\), vanishes for \(2<\alpha<3\), and correctly reproduces the behavior \(M_+ \sim M_0\beta/x\) in the logarithmic case \(m=1,\alpha\to4\). The entropy either vanishes (for \(\sigma>0\)), diverges (for \(\sigma<0\)), or diverges logarithmically (for \(\sigma=0\)).  Accordingly, the plus-sign branch does not provide the
same temperature regularization, as it admits no minimum
radius and therefore does not prevent the limit
$T_h\to\infty$. Only the minus-sign branch, characterized
by the natural cutoff $x_{\min}=\beta^{1/(\alpha-2)}$ and
a finite minimum entropy, provides a thermodynamically
well-defined and consistent framework.

Since quantum-gravity corrections to the Bekenstein
entropy can take either sign, the formalism admits two
distinct solution branches. Temperature regularization
nevertheless discriminates between them. The minus
branch of the MHR imposes a lower bound on the horizon
radius at $x_{\min}$, limits the Cai--Kim temperature to
\[
T_{\max}=\frac{\hbar c}{2\pi k_B\,x_{\min}\,\ell_p},
\]
and yields a nonvanishing residual entropy
$S^{(-)}_{G,\min}>0$ in the limit of vanishing mass. The
plus branch possesses no analogous cutoff, allowing
$x\to0$ and consequently $T_h\to\infty$, and therefore
does not satisfy this criterion. The combined requirements
of entropy positivity and temperature regularization thus
operate as a selection mechanism that favors the minus
branch when this criterion is imposed, thereby fixing the
sign of the entropy correction that would otherwise remain
ambiguous. 

We note, however, that the sign of the
quantum-gravity correction has broader implications
beyond temperature regularization: it was shown
in~\cite{Ong:2018syk} that a negative GUP parameter, which gives rise to a positive
logarithmic correction to the Bekenstein--Hawking entropy (which in
the MHR framework corresponds to the minus branch) allows
black holes to evaporate completely in infinite time,
leading to the formation of a metastable remnant. The
implications of this feature for black hole solutions
within the present framework require further
investigation. Nevertheless, both signs of the
quantum-gravity correction are retained in the
cosmological analysis below in order to determine whether
they produce distinguishable effects on the background
evolution.

In the specific case of an entanglement-induced correction
(with $\beta=1$, $m=1$, and $\alpha$ left as a free parameter), the cutoff reduces to $x_{\min}=1$, independently of $\alpha$. This limits the temperature to the Planck scale,
\[
T_{\max}
=
\frac{\hbar c}{2\pi k_B\ell_p}.
\]
The corresponding residual entropy,
\[
S^{(-)}_{G,\min}
=
\pi\gamma k_B\frac{\alpha-2}{4-\alpha},
\]
remains finite and positive for $2<\alpha<4$ on the minus branch,
whereas the plus branch entails no analogous cutoff.

}

{We note that the generalized entropy~\eqref{eq:entropy_def}, like the Bekenstein--Hawking entropy, diverges in the limit $T_h\to0$. In standard thermodynamics, as $T\to0$, Planck's formulation of the third law requires the entropy to approach a constant, conventionally taken to be zero. Israel showed that black-hole dynamics has no analogue of this formulation; instead, the third law is expressed through the unattainability of zero surface gravity in finite advanced time~\cite{Israel:1986gqz}. Thus, the formal divergence of the horizon entropy as $T_h\to0$ does not constitute a violation of the black-hole third law. In the present cosmological setting, $T_h\propto r_a^{-1}$,
so that $T_h\to0$ would require $r_a\to\infty$, which none of
the solutions reach within the cosmological range considered here.}

Following the procedure outlined in Sec. \ref{sec:First}, we substitute Eq.~\eqref{eq:entropy_def} into the Cai--Kim first law \eqref{fi}. This procedure yields the corresponding modified acceleration equation
\begin{equation}\label{eq:acceleration_equation_mod}
 m\gamma
\left[
\frac{r_a^{m-1}}{\ell_p^{m-1}}
\mp
\beta(\sigma-1)
\frac{r_a^{\sigma-2}}{\ell_p^{\sigma-2}}
\right]
\left(
\dot H-\frac{k c^2}{a^2}
\right)
=
-\frac{4\pi G}{c^2}
(\varepsilon+P),
\end{equation}
and using Eq.~\eqref{continuity equation} and integrating, we obtain the modified Friedmann equation
\begin{widetext}
\begin{equation}\label{eq:friedmann_mod}
2m\gamma c^2
\left[
\frac{1}{3-m}
\frac{
\left(\frac{H^2+\frac{k c^2}{a^2}}{c^2}\right)^{\frac{3-m}{2}}
}{\ell_p^{m-1}}
\mp
\beta
\frac{\sigma-1}{4-\sigma}
\frac{
\left(\frac{H^2+\frac{k c^2}{a^2}}{c^2}\right)^{\frac{4-\sigma}{2}}
}{\ell_p^{\sigma-2}}
\right]
=
\frac{8\pi G}{3c^2}\varepsilon
+\frac{\Lambda c^2}{3}.
\end{equation}
\end{widetext}
For a spatially flat universe, Eq.~\eqref{eq:friedmann_mod} can be written in terms of the standard Friedmann equation
\begin{equation}\label{dimensionless equation}
H^2=\frac{8\pi G}{3c^2}
(\varepsilon_m+\varepsilon_r+\varepsilon_{de}),
\end{equation}
by 
following Ref.~\cite{Lymperis:2018iuz}, where we have defined the effective dark-energy density 
\begin{multline}\label{ede}
\varepsilon_{de}
=
\frac{3c^2}{8\pi G}
\Bigg[
\frac{\Lambda c^2}{3}
+
H^2
\Bigg(
1
-
\frac{2m\gamma}
{(3-m)\ell_p^{m-1}}
\left(\frac{H}{c}\right)^{1-m}
\\
\pm
\frac{2m\gamma\beta(\sigma-1)}
{(4-\sigma)\ell_p^{\sigma-2}}
\left(\frac{H}{c}\right)^{2-\sigma}
\Bigg)
\Bigg].
\end{multline}
{It should be noted that the change of sign from $(\mp)$ in
Eq.~\eqref{eq:friedmann_mod} to $(\pm)$ in Eq.~\eqref{ede}
arises from transferring the correction term in Eq. \eqref{eq:friedmann_mod} to the effective
dark-energy sector. Accordingly, the upper $(-)$ and lower $(+)$
branches in Eqs.~\eqref{EMHR} and \eqref{eq:entropy_def}
correspond, respectively, to the upper $(+)$ and lower $(-)$ signs
in Eq.~\eqref{ede} and in the dimensionless expressions below.}

Since $\varepsilon_{de}$ depends explicitly on $H$, the cosmological evolution must be obtained numerically by solving the above equations simultaneously.  With the numerical solutions in hand, we assume that the effective dark-energy component satisfies the continuity equation with the equation of state $\omega_{de}$
\begin{equation}\label{decontinuity}
\omega_{de}
=
-1
-
\frac{\dot{\varepsilon}_{de}}
{3H\varepsilon_{de}}.
\end{equation}
The modified Friedmann equation and the corresponding effective dark energy density can be expressed in terms of the normalized Hubble parameter, $H(a)=H_0E(a)$, and the dimensionless density parameters $\Omega_i=\varepsilon_i/\varepsilon_c$, where $i=m,r,de$ denote the matter, radiation, and effective dark energy components, respectively and
%\begin{equation}   
$\varepsilon_c=\frac{3c^2H^2}{8\pi G}$
%\end{equation}
is the critical energy density. The resulting Friedmann equation takes the form
\begin{equation}\label{Ea}
E^2(a)=\frac{\Omega_{m0}a^{-3}+\Omega_{r0}a^{-4}}
{1-\Omega_{de}(a,E)},
\end{equation}
where
\begin{equation}\label{Omde_caseII}
\Omega_{de}(a,E)=\frac{\Omega_{\Lambda0}}{E^2}+1-AE^{1-m}\pm BE^{2-\sigma}.
\end{equation}
The present-day density parameters are defined by
%\begin{equation}    
$\Omega_{i0}=\frac{8\pi G}{3c^2H_0^2}\varepsilon_{i0}$,
%\end{equation}
with $i=m,r,de$. The dimensionless coefficients $A$ and $B$ are given by
\begin{equation}\label{AB}
A=\frac{2m\gamma}{3-m}\left(\ell_p\frac{H_0}{c}\right)^{1-m},
~
B=\frac{2m\gamma\beta(\sigma-1)}{4-\sigma}
\left(\ell_p\frac{H_0}{c}\right)^{2-\sigma}.
\end{equation}
These coefficients parameterize the contributions arising from the leading and subleading terms in the generalized entropy. In particular, $A$ depends on the parameters governing the leading-order modification, whereas $B$ quantifies the contribution of the subleading correction and vanishes in the limit $\beta=0$.

Imposing the flatness condition at the present epoch $(a=1)$ determines the present value of the cosmological constant,
\begin{equation}\label{OmLambda}
\Lambda=\frac{3H_0^2}{c^2}
\left[A\mp B-(\Omega_{m0}+\Omega_{r0})\right].
\end{equation}
In terms of these dimensionless quantities, the effective dark-energy density \eqref{ede} can be written as
\begin{equation}\label{ede_compact}
\varepsilon_{de}=
\varepsilon_{c0}\Omega_{\Lambda0}
+\left(1-AE^{1-m}\pm BE^{2-\sigma}\right)\varepsilon_c,
\end{equation}
with $\omega_{de}$ given by Eq.~\eqref{decontinuity}. The deceleration parameter is
\begin{equation}
q=-1-\frac{aE'}{E}.
\end{equation}

Since $\Omega_{de}$ depends explicitly on $E$, the cosmological evolution must be determined numerically. Arbitrary-precision arithmetic\footnote{The numerical analysis employs the arbitrary-precision Python library \texttt{mpmath} \cite{mpmath} with 50-digit precision, using SI units throughout. Equations~\eqref{Ea} and \eqref{Omde_caseII} are solved at each value of the scale factor through a two-dimensional Newton--Raphson iteration \cite{Press:2007ipz} over the interval (a $\in[10^{-10}$,1]). This range is required because modified cosmological models based on generalized entropies may remain observationally viable at low redshift ($z\lesssim2$) while exhibiting pathological behavior at earlier times, including negative effective dark-energy densities or divergences in the equation of state \cite{Ibarbo-Perlaza:2025hwi}.} is employed to ensure numerical stability when the entropy corrections involve powers of the extremely small quantity $\ell_p H_0/c$.

To investigate the cosmological implications of the model, we consider specific parameter choices corresponding to physically motivated realizations of the generalized mass-to-horizon entropy. For each case, we present the entropy $S_G$ and the corresponding effective dark-energy density parameter $\Omega_{de}(a,E)$, which serve as inputs to the numerical analysis.

\subsection{Standard Area-Law Extensions (\texorpdfstring{$\beta = 0$}{beta = 0})}
\label{sec:caseI}

Imposing $\beta = 0$ reduces the generalized entropy to
\begin{equation}
S_G=\frac{2\pi\gamma k_B m}{m+1}
\left(\frac{r_a}{\ell_p}\right)^{m+1}.
\end{equation}

This family of models encompasses several notable limiting cases: the Bekenstein--Hawking entropy for $m = \gamma = 1$, the Tsallis--Cirto entropy for $m = 2\delta - 1$ \cite{Tsallis:2012js}, and the Barrow entropy for $m = 1 + \Delta$ \cite{Barrow:2020tzx}. Within this class, the associated dark-energy density parameter takes the form
\begin{equation}
\Omega_{de}(a,E)=
\frac{\Omega_{\Lambda0}}{E^2}
+1-AE^{1-m}.
\end{equation}

\subsection{Quantum-Entanglement Corrections}\label{sec:caseII}

For $\beta = 1$ and $2 < \alpha < m + 2$, the generalized entropy acquires a subleading power-law correction that can be attributed to quantum entanglement across the horizon \cite{Das:2007mj,Sheykhi:2010yq},
\begin{equation}
\resizebox{0.98\columnwidth}{!}{$
\displaystyle
S_G=2\pi\gamma k_B
\left[
\frac{m}{m+1}
\left(\frac{r_a}{\ell_p}\right)^{m+1}
\mp
m\frac{m+2-\alpha}{m+3-\alpha}
\left(\frac{r_a}{\ell_p}\right)^{m+3-\alpha}
\right]
$}
\end{equation}
{Although the correction remains subleading for all
$\alpha>2$, we restrict the power-law corrections to
$2<\alpha<m+2$ so that $(\sigma-1)/\sigma>0$ and the upper
$(-)$ branch retains the negative sign of the power-law
entanglement correction
\cite{Sheykhi:2010yq,Komatsu:2017gtf}.}

In this case, the dark-energy density parameter is modified to
\begin{equation}
\Omega_{de}(a,E)=
\frac{\Omega_{\Lambda0}}{E^2}
+1-AE^{1-m}
\pm BE^{\alpha-m-1}.
\end{equation}

This construction therefore includes, as particular instances, the entanglement-corrected generalizations of the Bekenstein--Hawking, Tsallis--Cirto, and Barrow entropies.
\subsection{Quantum-gravity corrections}\label{sec:caseIII}

By fixing $\alpha = 4$, one obtains quantum-gravity-induced corrections to the entropy \cite{Gohar:2025yfx,Gohar:2026hiy} of the form
\begin{equation}
S_G =
\frac{2\pi\gamma k_B m}{m+1}
\left(\frac{r_a}{\ell_p}\right)^{m+1}
\mp
2\pi\gamma k_B m\beta
\frac{m-2}{m-1}
\left(\frac{r_a}{\ell_p}\right)^{m-1}.
\end{equation}

The corresponding dark-energy density parameter is then modified to
\begin{equation}
\Omega_{de}(a,E) =
\frac{\Omega_{\Lambda0}}{E^2}
+ 1 - A E^{1-m}
\pm B E^{3-m}.
\end{equation}

In the limit $m \rightarrow 1$ with $\gamma = 1$, the entropy expression converges to the standard logarithmically corrected Bekenstein-Hawking entropy,
\begin{equation}\label{S_caseIII}
S_G =
\pi k_B\left(\frac{r_a}{\ell_p}\right)^2
\pm
2\pi k_B \beta
\ln\left(\frac{r_a}{\ell_p}\right)
+\mathrm{const},
\end{equation}
thereby reproducing the well-known microcanonical and canonical quantum-gravity corrections derived in various approaches \cite{Rovelli:1996dv,Medved:2004yu,Medved:2004eh}.
\section{Results And Discussion}
\label{Discussion}
The background evolution is obtained by numerically solving the dimensionless
Friedmann equation~\eqref{Ea} simultaneously with Eq.~\eqref{Omde_caseII}.
Results for the three physically motivated parameter configurations of the
generalized mass-to-horizon entropy~\eqref{eq:entropy_def} are presented in
Figs.~\ref{fig:caseI} and~\ref{fig:caseI_gamma}. In each figure, the left panel displays the energy densities $\varepsilon_i(a)$ and the right panel the effective equation of state $\omega_{de}(a)$. The fiducial cosmological parameters are fixed to
$\Omega_{m0}=0.31$, $\Omega_{r0}=9\times10^{-5}$, and $H_0=73~\mathrm{km\,s^{-1}\,Mpc^{-1}}$, with the cosmological constant
determined self-consistently from Eq.~\eqref{OmLambda} for each model. The
$\Lambda$CDM limit $(m=\gamma=1,\,\beta=0)$ is indicated by the solid blue curve
throughout. The effective dark-energy density~\eqref{ede_compact} decomposes into
a constant contribution $\varepsilon_{c0}\Omega_{\Lambda0}$ and a dynamical
correction proportional to $\varepsilon_c(a)=\varepsilon_{c0}E^2(a)$. Depending
on the powers of $E(a)$ and the coefficients $A$ and $B$, this correction can
grow substantially or become negative at early times. To preserve the standard
radiation--matter--dark-energy sequence and ensure observational consistency, following Ref.~\cite{Ibarbo-Perlaza:2025hwi}, we
impose $\Omega_{de}<0.045$ at Big Bang Nucleosynthesis~\cite{Bean:2001wt} and matter--radiation
equality, alongside
$\Omega_{de}<0.02$ at $a\simeq0.02$ from CMB constraints~\cite{Planck:2015bue}. 

\subsection{Standard Area-Law Extensions (\texorpdfstring{$\beta = 0$}{beta = 0})}
\label{secLg}

We first consider the standard area-law extension discussed in Sec.~\ref{sec:caseI}, characterized by the entropy exponent \(m\) and the MHR coupling parameter \(\gamma\). The corresponding entropy and effective dark-energy density parameter were derived in Sec.~\ref{sec:caseI}, while the effective dark-energy density is given by Eq.~\eqref{ede_compact}. The cosmological dynamics are governed by the correction term ($AE^{1-m}$), whose amplitude is determined by \(m\) and \(\gamma\). The resulting cosmological evolution is presented in Figs.~\ref{fig:caseI} and \ref{fig:caseI_gamma}.

\subsubsection{Barrow and Tsallis-Cirto type entropy functions (\texorpdfstring{$m$}{m},\texorpdfstring{$\gamma$}{gamma})}
\label{secBTCmg}

\begin{figure*}[!t]
\caption{Background evolution of the standard area-law extension ($\beta = 0$) at fixed $\gamma = 1$ (top panels) and for varying $\gamma$ (bottom panels). Left panels: energy density. Right panels: effective dark energy equation of state. Top panels: for $m = 1.0001$ at $\gamma = 1$, $\varepsilon_{de}$ is negative for $a \lesssim 10^{-1}$ (top-left inset), and its sign change near $a = 10^{-1}$ produces the divergence in $\omega_{de}$ (top right). Varying $\gamma$ jointly with $m$ ($\gamma = 0.98$) keeps $\varepsilon_{de}$ positive (top left) and returns $\omega_{de}$ to the quintessence regime (top right). Bottom panels: adjusting $\gamma$ alongside $m$ keeps $\varepsilon_{de}$ positive at all epochs (bottom left). For $\gamma = 0.8705$ and $0.871$, $\varepsilon_{de}$ is non-monotonic and passes through a minimum near $a = 10^{-1}$ (bottom-left inset), where $\omega_{de}$ crosses the phantom divide $\omega_{de} = -1$ (bottom right).}
\centering
\includegraphics[width=0.99\linewidth]{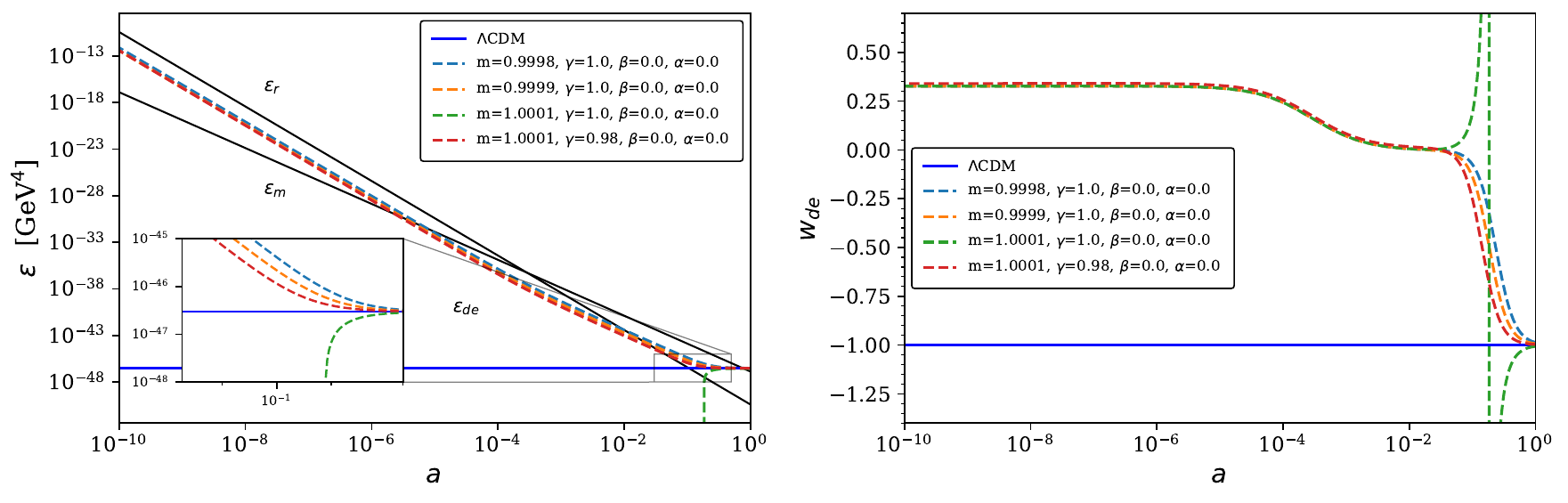}\\
~~~\\
\includegraphics[width=0.99\linewidth]{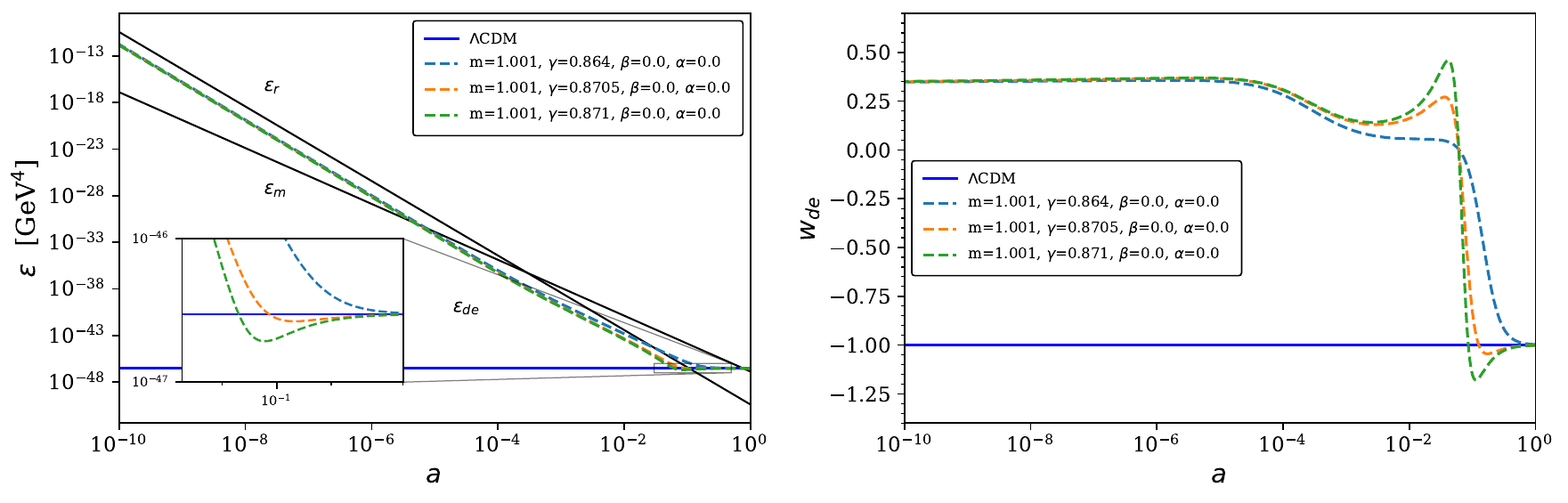}
\label{fig:caseI}
\end{figure*}

\begin{figure*}[!t]
\caption{Background evolution for the $\gamma$-scaled Bekenstein-Hawking entropy with $m = 1$, $\beta = 0$ (top panels), for the power-law entanglement corrections with $\beta = 1$, $2 < \alpha < m+2$ (middle panels), and for the QG corrections with $\alpha = 4$, $\beta = 0.1$ (bottom panels). Left panels: energy density. Right panels: effective dark energy equation of state. Top panels: for $m = 1$ the correction term is independent of $E$, so $\varepsilon_{de}$ carries a constant fraction $(1-\gamma)$ of the critical density and tracks the dominant background component through the radiation and matter eras (top left). The sub-scaled cases ($\gamma < 1$) give a quintessence $\omega_{de}$ (top right), while $\gamma = 1.001$ gives a negative $\varepsilon_{de}$ (top left). Middle panels: the correction term $BE^{\alpha-m-1}$ enhances $\varepsilon_{de}$ at small scale factors and decays toward late times (middle left). For $m = \gamma = 1$, $\alpha = 2.95$ is indistinguishable from $\Lambda$CDM, while $\alpha = 2.04$ gives a moderate early-time contribution; the case $m = 1.001$, $\gamma = 0.872$, $\alpha = 2.05$ (purple dashed) has a larger early-time contribution from $m > 1$ and a late-time evolution close to $\Lambda$CDM. Bottom panels: for $m = \gamma = 1$ both the canonical $(+)$ and microcanonical $(-)$ branches coincide with $\Lambda$CDM, since the prefactor is suppressed to $B \sim 10^{-122}$; the deviations for $(1, 0.985)$ and $(1.001, 0.871)$ arise from the leading area-law term rather than from the QG correction (left and right panels).}
\centering
\includegraphics[width=0.99\linewidth]{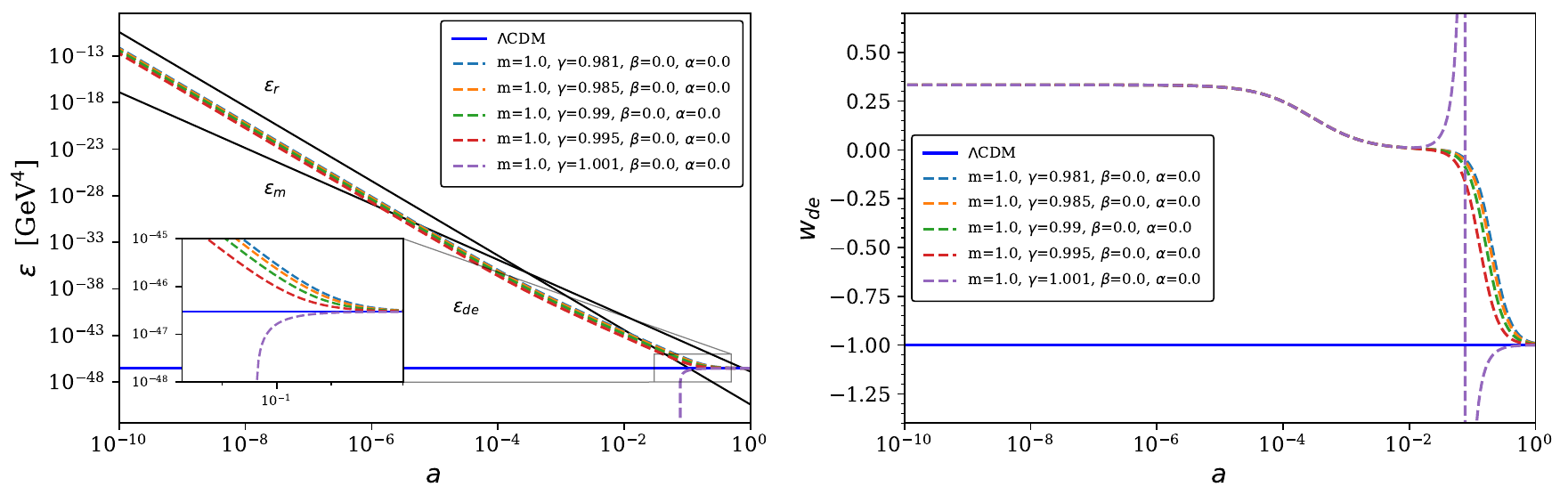}\\
~~~\\
\includegraphics[width=\linewidth]{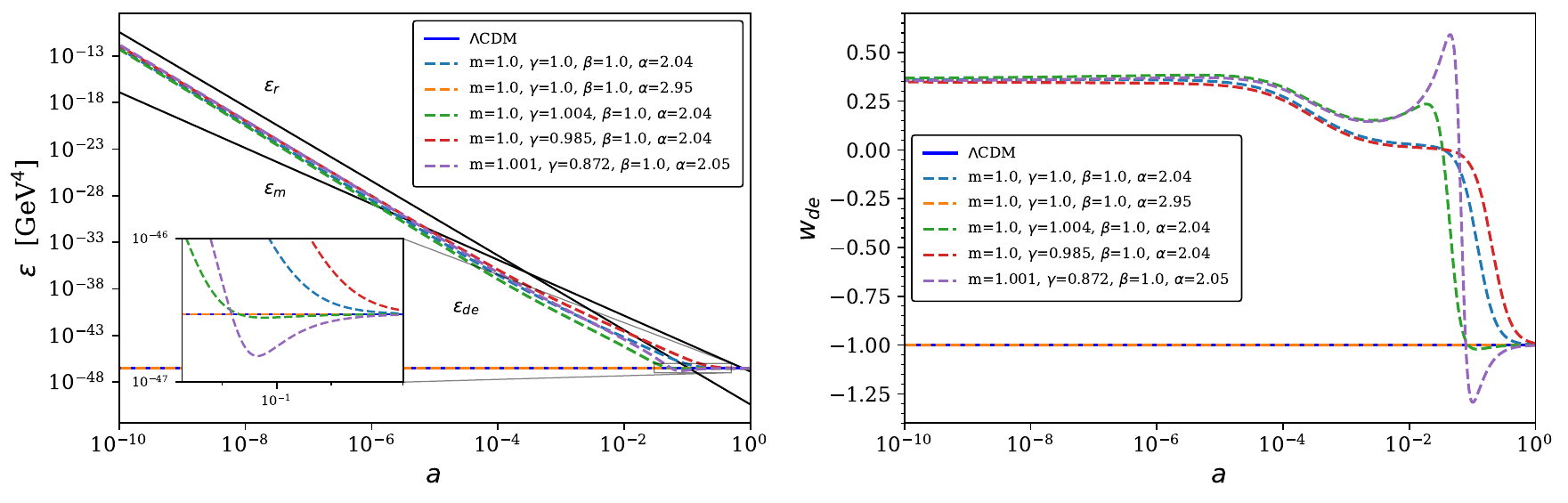}\\
~~~\\
\includegraphics[width=\linewidth]{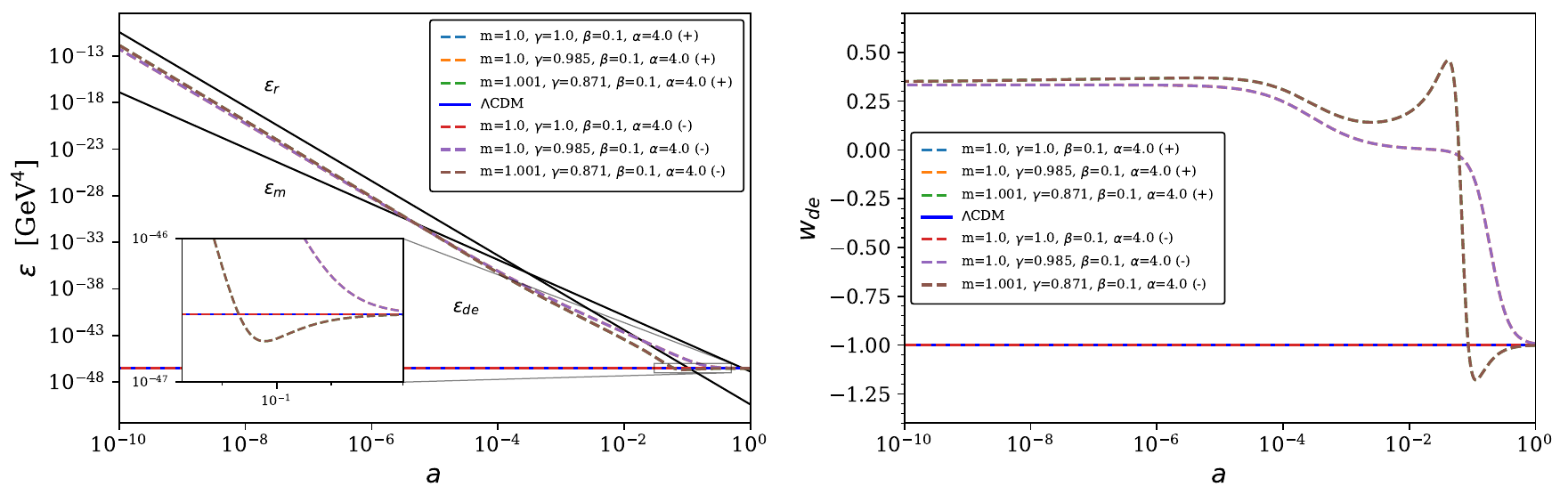}
\label{fig:caseI_gamma}
\end{figure*}

We first fix $\gamma=1$ and vary $m$, allowing the entropy to scale as
$S_G\propto r_a^{m+1}$. This encompasses the Tsallis--Cirto $(m=2\delta-1)$ and
Barrow $(m=1+\Delta)$ entropies. In contrast to previous approaches, the
generalized MHR~\eqref{EMHR} ensures that the horizon energy scales consistently
with the modified entropy, $M\propto L^m$.

The coefficient $A$ contains the factor $(\ell_pH_0/c)^{1-m}\approx10^{-61(1-m)}$.
Consequently, even infinitesimal deviations from $m=1$ produce large changes in
$A$. For $m>1$, $AE^{1-m}>1$, rendering $1-AE^{1-m}$ negative; for $m<1$, it
remains positive. This sensitivity is evident in Fig.~\ref{fig:caseI}.

For $m<1$, the effective dark-energy density remains positive and tracks the
dominant background component during the radiation and matter eras. Once the
constant term $\varepsilon_{c0}\Omega_{\Lambda0}$ dominates, the model enters a
phase of accelerated expansion. The case $m=0.9999$ satisfies all observational
constraints and exhibits a quintessence-like equation of state. {However, analysis shows that even the
marginal deviation $m=0.9998$ violates the CMB bound, yielding
$\Omega_{de}(a\simeq0.02)=0.027$.}

For $m>1$, the correction term becomes negative, driving both $\Omega_{de}$ and
$\varepsilon_{de}$ below zero at early times. The flatness condition then forces
$\Omega_r$ and $\Omega_m$ to exceed unity, while the zero-crossing of
$\varepsilon_{de}$ produces the divergence in $\omega_{de}$ visible in
Fig.~\ref{fig:caseI}. For $m=1.0001$, these effects already manifest at the
$10^{-4}$ level.

The pathology intensifies with increasing $|m-1|$: {$m>1$ drives the dark-energy density deeper into negative values, while $m<1$ generates an excessive dark-energy fraction during the radiation and matter eras. Accordingly, at fixed $\gamma=1$, viability is restricted to $0.99986\le m\le1$.}

Notably, allowing $\gamma$ to vary alleviates these pathologies. {Configurations excluded at $\gamma=1$ such as}
$(m,\gamma)=(0.9998,1)$ and $(1.0001,1)$, become viable upon adjustment to
$(0.9998,1.01)$ and $(1.0001,0.98)$, respectively. The corresponding evolution is
illustrated in Fig.~\ref{fig:caseI_gamma}. This demonstrates that departures from
the standard area law cannot be introduced through $m$ alone; a simultaneous
adjustment of $\gamma$ is required to maintain a physically consistent dark-energy
sector.

We next examine whether this compensation mechanism remains effective for larger
deviations, $|m-1|\sim10^{-3}$.

The interplay between $m$ and $\gamma$ is illustrated in Fig.~\ref{fig:caseI_gamma}, where
for $m=1.001$ {we identify a viable interval $0.864\le\gamma\le0.871$. Within this range,} the
$\gamma$-adjustment eliminates the negative-density pathology present at $\gamma=1$,
keeping $\Omega_{de}$ and $\varepsilon_{de}$ positive and satisfying the
early-universe bounds.

{At the lower end of the interval}, $\varepsilon_{de}$ evolves nearly parallel to the dominant
background fluid and $\Omega_{de}$ remains close to the BBN bound with $\Omega_{de}^{\rm BBN}=0.044$. The equation of state remains in the quintessence regime, yielding  $\omega_{de}^{\text{today}} = -0.997$. {Towards the upper end of the interval ($\gamma\simeq0.871$)}, $\varepsilon_{de}$ develops a shallow minimum near
$a\sim10^{-1}$, producing a transient increase in $\omega_{de}$ followed by a
crossing of the phantom divide. The present-day equation of state becomes
marginally phantom,  $\omega_{de}^{\text{today}}=-1.001$. { In both regimes, $q_0$ remains close to its $\Lambda$CDM Value}. The evolution of $\varepsilon_{de}$ therefore
determines whether the model behaves as quintessence or undergoes a phantom-divide
crossing.

{The branch $m<1$ admits no viable value of $\gamma$: increasing $\gamma$ reduces the early dark-energy contribution but cannot simultaneously satisfy the positivity and CMB constraints, so this branch is excluded.}

Larger deviations on the $m>1$ side are similarly problematic. For {instance, with $m=1.01$,}
adjusting $\gamma$ eliminates the negative-density pathology but generates
excessive early dark energy. Although the present-day quantities remain consistent
with $\Lambda$CDM, the BBN and CMB bounds are violated, and no viable solution
exists for $|m-1|\sim10^{-2}$.

Overall, varying
$\gamma$ extends viability only modestly, from $|m-1|\sim10^{-4}$ to
$|m-1|\sim10^{-3}$ on the $m>1$ side. Deviations of order $10^{-2}$ remain
excluded; we therefore adopt $m=1.001$ in the subsequent analysis.

A related model was studied in Ref.~\cite{Basilakos:2025wwu}, where their
dark-energy term, recast in the dimensionless form employed here, gives the
coefficient $A^{n\gamma}=\frac{2n\gamma}{3-n}H_0^{1-n}$,
evaluated in units $\hbar=c=k_B=8\pi G=1$ with $H_0=1$.\footnote{{Ref.~\cite{Basilakos:2025wwu} employs the MHR $M=\gamma\frac{c^2}{G}r_a^n$ ~\cite{Gohar:2023hnb}, in which  the coupling parameter $\gamma$ carries dimensions $\text{[L]}^{1-n}$, whereas in the generalized MHR \eqref{EMHR} adopted here, $\gamma$ is dimensionless.}} Restoring the Planck
scale yields $H_0\sim10^{-61}$, which drives $A^{n\gamma}$ far from unity for
even small departures from $n=1$, so the same pathologies identified above emerge
when the evolution is extended to high redshift. In the present framework, where
the Planck length is retained explicitly in $A$ [Eq.~\eqref{AB}]
the resulting parameter sensitivity constrains both the Tsallis exponent $\delta$
and the Barrow exponent $\Delta$ to remain extremely close to their standard
values.
\subsubsection{\texorpdfstring{$\gamma$}{gamma}-rescaled Bekenstein-Hawking entropy (\texorpdfstring{$m=1$}{m=1})}
\label{gammafreesb}
We now fix $m=1$ and allow the MHR coupling parameter $\gamma$ to vary. In this case, the entropy preserves the standard area-law scaling,
$S_G=\gamma S_{\rm BH}$,
while the modified term $AE^{1-m}$ reduces to $A=\gamma$. As a consequence, the correction entering $\Omega_{de}$ becomes independent of $E(a)$ and manifests as a constant offset with respect to the $\Lambda$CDM model.

For $a\lesssim10^{-1}$, the deviation $(1-\gamma)$ contributes a constant fraction of the total energy density, such that $\varepsilon_{de}$ evolves approximately as $(1-\gamma)\varepsilon_c(a)$ and tracks the dominant background component. This behavior is illustrated in Fig.~\ref{fig:caseI_gamma}. At late times, the cosmological-constant contribution becomes dominant and drives accelerated expansion. 

The associated equation-of-state parameter evolves from radiation-like behavior at very early times to matter-like behavior around $a\sim10^{-2}$, and subsequently asymptotes to $-1$ at late times. The deceleration parameter remains close to its $\Lambda$CDM value, with only modest shifts in $q_0$. {The early-universe  contribution remains essentially constant across the epochs considered.} The choice $\gamma=0.981$ nearly saturates the CMB constraint, whereas $\gamma=0.980$ yields $\Omega_{de}(a\simeq0.02)=0.020$, delineating the boundary of the allowed parameter space.

The sign of the deviation is crucial. Sub-scaled entropies ($\gamma<1$) generate positive dark-energy densities and lead to viable, quintessence-like cosmologies. In contrast, super-scaled entropies ($\gamma>1$) give rise to negative $\Omega_{de}$ and $\varepsilon_{de}$, reproducing the pathologies discussed above. The observationally viable range of the coupling parameter is therefore constrained to
$0.981 \le \gamma \le 1.$

\subsection{Quantum Entanglement Corrections}

We now analyze the power-law entanglement corrections to the generalized mass-to-horizon entropy~\eqref{eq:entropy_def} introduced in Sec.~\ref{sec:caseII}. Since these corrections are predominantly governed by the exponent $\alpha$, we fix $\beta=1$ and study the cosmological dynamics as a function of $\alpha$, together with $\gamma$ and $m$. The corresponding results are presented in Fig.~\ref{fig:caseI_gamma}. {Throughout this section, we adopt the upper sign in
Eq.~\eqref{Omde_caseII}, which corresponds to the upper $(-)$ branch
of the MHR and entropy. For this choice,} the entanglement term yields a positive contribution to the effective energy density and increases toward smaller scale factors via the factor $BE^{\alpha-m-1}$.

\subsubsection{Entanglement corrections with (\texorpdfstring{$m=1$}{m=1})}

We first focus on the case $m=\gamma=1$, which corresponds to entanglement-induced corrections to the standard Bekenstein–Hawking entropy~\cite{Das:2007mj,Sheykhi:2010yq}. In this setting, the allowed parameter space is restricted to $2<\alpha<3$, and the background evolution is entirely determined by the entanglement exponent $\alpha$.

In the vicinity of the upper bound, $\alpha\rightarrow3$, the coefficient $B\propto2(3-\alpha)/\alpha$ is strongly suppressed. For instance, at $\alpha=2.95$, one finds $B\approx4.8\times10^{-60}$, rendering the entanglement correction negligible. The resulting cosmological evolution is then indistinguishable from that of the $\Lambda$CDM model, as illustrated in Fig.~\ref{fig:caseI_gamma}. The upper bound of the admissible range for $\alpha$ thus corresponds to a weak-entanglement regime that effectively reproduces a cosmological-constant–dominated universe, in agreement with Ref.~\cite{Komatsu:2017gtf}.

At the opposite end of the parameter range, for $\alpha=2.01$, {we find that} the coefficient $B\approx0.24$ and the entanglement correction remains non-negligible throughout the entire expansion history. In this regime, the effective dark-energy density becomes comparable to the dominant background component during both radiation and matter domination, with $\Omega_{de}^{\rm BBN}=0.368$, $\Omega_{de}^{\rm eq}=0.274$, and $\Omega_{de}^{a\simeq0.02}=0.256$, all of which exceed current observational bounds. Because the correction only decays slowly, it also continues to influence late-time cosmology, giving $\omega_{de}^{\rm today}=-0.855$ and $q_0=-0.385$. The lower bound on $\alpha$ therefore characterizes a strong-entanglement regime that is ruled out by observations.

An intermediate choice, $\alpha=2.04$, yields a phenomenologically viable scenario. In this case, the entanglement contribution is still present but decays sufficiently rapidly that the cosmological-constant term dominates at late epochs. As shown in Fig.~\ref{fig:caseI_gamma}, the effective dark-energy density $\varepsilon_{de}$ initially tracks the dominant background fluid and subsequently approaches a cosmological-constant–like behavior, resulting in $\omega_{de}^{\rm today}=-0.998$ and $q_0=-0.533$. We therefore adopt $\alpha=2.04$ as the reference value in the remainder of our analysis.

We next vary $\gamma$ while holding $\alpha=2.04$ fixed. In this case, the entanglement contribution is modified by the constant {offset $(1-\gamma)$: values } $\gamma<1$ enhance the early dark-energy fraction, whereas $\gamma>1$ suppress it. For example, $\Omega_{de}^{\rm BBN}$ increases from $0.018$ at $\gamma=1$ to $0.032$ for $\gamma=0.985$, while it decreases to $0.014$ for $\gamma=1.004$.

The late-time dynamics reflect the relative strength of the correction. The stronger correction realized for $\gamma=0.985$ keeps the model in the quintessence regime with $\omega_{de}^{\rm today}=-0.991$, whereas the weaker correction for $\gamma=1.004$ leads to the non-monotonic behavior discussed in Sec.~\ref{secBTCmg}, including a marginal crossing of the phantom divide and $\omega_{de}^{\rm today}=-1.0002$. Consequently, stronger entanglement corrections drive the expansion farther away from the $\Lambda$CDM behavior, while weaker entanglement corrections tend to bring it closer.

\subsubsection{Entanglement corrections with (\texorpdfstring{$m\neq1$}{m neq 1})}

We now extend the analysis beyond the standard area law by activating the viable non-extensive solution $(m=1.001)$ identified in Sec.~\ref{secBTCmg}. 

{For $(m,\gamma,\alpha)=(1.001,0.874,2.04)$, we find a mild violation of the BBN constraint, with $\Omega_{de}^{\rm BBN}=0.051$. Increasing the entanglement exponent to $\alpha=2.05$ restores compatibility with BBN, reducing the early dark-energy fraction to $\Omega_{de}^{\rm BBN}=0.042$.} The corresponding cosmological evolution is displayed in Fig.~\ref{fig:caseI_gamma}. The dark-energy density remains positive over the entire expansion history and approaches, but does not exceed, the BBN limit at early times. As discussed in Sec.~\ref{secBTCmg}, the non-monotonic evolution of $\varepsilon_{de}$ induces a crossing of the phantom divide, resulting in a slightly phantom present-day equation-of-state parameter, $\omega_{de}^{\rm today}=-1.002$, with a deceleration parameter $q_0=-0.536$.

In summary, the entanglement-induced contribution is jointly governed by $\alpha$, $\gamma$, and $m$. The exponent $\alpha$ determines the intrinsic amplitude of the correction, whereas $\gamma$ and $m>1$ enhance its relative importance at early times. For all phenomenologically admissible parameter combinations, the correction term increases toward small scale factors and decays at late times, in agreement with the behavior reported in Ref.~\cite{Sheykhi:2010yq}.

\subsection{QG Corrections (\texorpdfstring{$\alpha =4$}{alpha = 4})}

We now examine the quantum-gravity–corrected sector introduced in Sec.~\ref{sec:caseIII}, corresponding to $\alpha=4$ (equivalently, $\sigma=m-1$). The entropy expression~\eqref{S_caseIII} includes a quantum-gravity (QG) correction whose sign distinguishes the canonical $(+)$ from the microcanonical $(-)$ branch. Since this correction is predominantly governed by the parameter $\beta$, we fix $\beta=0.1$ and vary $m$ and $\gamma$. The resulting cosmological dynamics are displayed in Fig.~\ref{fig:caseI_gamma}.

For $m=\gamma=1$, the leading correction term $1-AE^{1-m}$ vanishes identically, and Eq.~\eqref{AB} implies $B\propto(\ell_pH_0/c)^2\sim10^{-122}$. Accordingly, the QG contribution is completely negligible, independent of the value of $\beta$ and of the choice of canonical or microcanonical branch. The background evolution is thus exactly that of $\Lambda$CDM, as corroborated by Fig.~\ref{fig:caseI_gamma}. In this regime, the effective dark-energy density is strictly constant, $\varepsilon_{de}=\varepsilon_{c0}\Omega_{\Lambda0}$, and the dark-energy equation-of-state parameter remains fixed at $\omega_{de}=-1$ throughout cosmic history.

When either $\gamma\neq1$ or $m\neq1$, the cosmological evolution continues to be dominated by the term $AE^{1-m}$, while the genuine QG contribution remains suppressed by the Planck-scale factor $10^{-122}$. Consequently, the deviations from $\Lambda$CDM, illustrated by the dashed curves in Fig.~\ref{fig:caseI_gamma}, are effectively indistinguishable from those analyzed previously in Secs.~\ref{gammafreesb} and~\ref{secBTCmg}. In particular, the parameter choices \((1,0.985,0.1,4)\) and $(1.001,0.871,0.1,4)$ reproduce, respectively, the same quintessence-like and marginally phantom behaviors identified in those earlier sections.

Although the QG correction induces a formal structure analogous to that of the entanglement sector, its phenomenological impact is negligible because observationally admissible values of $m$ remain extremely close to unity and the coefficient $B$ is suppressed at the Planck scale. As a result, neither the canonical nor the microcanonical branch gives rise to observable modifications of the dark-energy sector. Within the present framework, the cosmological dynamics are dictated almost entirely by the MHR coupling parameter $\gamma$ and its interplay with the entropy exponent $m$, while the QG correction itself remains observationally irrelevant.

\section{Conclusions}
\label{conclusion}

We have examined the cosmological consequences of the generalized four-parameter mass-to-horizon entropy~\eqref{eq:entropy_def} within the Cai–Kim thermodynamic framework. The corresponding modified Friedmann equations depend on the entropy exponent $m$, the MHR coupling parameter $\gamma$, and the quantum-correction parameters $\beta$ and $\alpha$. The combined effect of these parameters can be recast as an effective dark-energy contribution. By solving the background equations for scale factors in the range $a\in[10^{-10},1]$, we constrained the parameter space using bounds on the dark-energy density parameter $\Omega_{de}$ derived from Big Bang Nucleosynthesis, matter–radiation equality, and Cosmic Microwave Background observations.

In the case of the standard area-law extension $(\beta=0)$, the cosmological dynamics exhibit a pronounced sensitivity to deviations from the Bekenstein–Hawking entropy. {For fixed $\gamma=1$, the entropy exponent is confined to a narrow interval around unity ($0.99986\le m\le1$), beyond which even small departures lead to either negative or excessively large effective dark-energy densities, thereby disrupting the standard radiation--matter--dark-energy sequence.}

For the $\gamma$-rescaled Bekenstein–Hawking entropy $(m=1)$, observational consistency is achieved only in the sub-scaled regime ($0.981\le\gamma\le1$), within which the effective dark-energy sector is positive and displays a quintessence-like behavior. The super-scaled regime $(\gamma>1)$ produces negative dark-energy densities and is consequently ruled out.

Entanglement-induced corrections $(\beta=1$, $2<\alpha<m+2)$ generate an early-time dark-energy contribution that naturally decays toward late times. Values of $\alpha$ near the lower bound give rise to an excessively large early dark-energy fraction and are excluded, whereas values close to the upper bound effectively reproduce $\Lambda$CDM. Admissible solutions therefore require only moderate entanglement corrections and remain close to a cosmological-constant behavior at the present epoch. By contrast, the quantum-gravity corrections controlled by $\beta$ are suppressed by the Planck-scale factor $(\ell_p H_0/c)^{3-m}$ and are entirely negligible on cosmological scales, irrespective of the underlying statistical origin.

Overall, our analysis confines the entropy exponent and MHR coupling parameter to a narrow region around the standard area-law values. Non-extensive Tsallis–Cirto and Barrow-type generalizations are thus tightly constrained, with the viable parameter space restricted to values close to the Bekenstein–Hawking entropy $(\delta\simeq1$, $\Delta\simeq0)$. For all viable models, the present-day dark-energy equation-of-state parameter remains close to $\omega_{de}=-1$, and the deceleration parameter is nearly indistinguishable from its $\Lambda$CDM counterpart.

These conclusions are based exclusively on background evolution and early-universe constraints. A comprehensive statistical analysis incorporating cosmological data to obtain quantitative parameter bounds will be presented in a forthcoming companion work.
\bibliographystyle{apsrev4-1}
\bibliography{references}

\end{document}